\documentclass[prl,twocolumn,amsmath,aps,superscriptaddress]{revtex4-1}

\usepackage{amsmath}
\usepackage{amssymb}
\usepackage{graphicx}
\usepackage{soul}
\usepackage{braket}
\usepackage{xcolor}

\begin{document}
    
    \title{Theory of Superconductivity at the LaAlO$_3$/SrTiO$_3$ heterointerface: Electron pairing mediated by deformation of ferroelastic domain walls}
    
    \author{David Pekker}
    \affiliation{Department of Physics and Astronomy, University of Pittsburgh, Pittsburgh, Pennsylvania 15260, USA}
    \affiliation{Pittsburgh Quantum Institute, Pittsburgh, Pennsylvania 15260, USA}
    \author{C. Stephen Hellberg}
    \affiliation{U.S. Naval Research Laboratory, Washington, DC 20375, USA}
    \author{Jeremy Levy}
    \affiliation{Department of Physics and Astronomy, University of Pittsburgh, Pittsburgh, Pennsylvania 15260, USA}
    \affiliation{Pittsburgh Quantum Institute, Pittsburgh, Pennsylvania 15260, USA}

\begin{abstract}
    SrTiO$_3$ is a superconducting semiconductor with a pairing mechanism that is not well understood.  SrTiO$_3$ undergoes a ferroelastic transition at $T=$ 105 K, leading to the formation of domains with boundaries that can couple to electronic properties.  At two-dimensional SrTiO$_3$ interfaces, the orientation of these ferroelastic domains is known to couple to the electron density, leading to electron-rich regions that favor out-of-plane distortions and electron-poor regions that favor in-plane distortion. Here we show that ferroelastic domain walls support low energy excitations that are analogous to capillary waves at the interface of two fluids. We propose that these capillary waves mediate electron pairing at the LaAlO$_3$/SrTiO$_3$ interface, resulting in superconductivity around the edges of electron-rich regions.  This mechanism is consistent with recent experimental results reported by Pai et al. [PRL $\bf{120}$, 147001 (2018)]
\end{abstract}
    
\maketitle
    
The origin of electron pairing in SrTiO$_3$ (STO) has remained mysterious for over half a century.  Superconductivity in bulk STO, first reported~\cite{Schooley1964} in 1964, takes place at exceedingly low carrier densities ranging from $8.5 \times 10^{18}$ to $3.0 \times 10^{20}\,\text{cm}^{-3}$ (Refs.~\cite{Koonce1967, Lin2013, Eagles2016, Swartz2018, Eagles1986}).  The superconducting transition temperature is a dome-shaped~\cite{Koonce1967} function of carrier density, reaching a maximum of $0.4~\text{K}$. Over the last five decades, there have been many attempts to identify the electron pairing mechanism, invoking in many cases the unusual or unique properties of STO.  Candidates for the pairing ``glue" have included valley degeneracy~\cite{Schooley1964}, longitudinal optical phonons~\cite{Cohen1964, Baratoff1981, Gorkov2016, Klimin2014}, antiferrodistortive modes~\cite{Appel1969}, plasmons~\cite{Ruhman2016}, plasmons in conjunction with optical phonons~\cite{Takada1980}, Jahn-Teller bipolarons~\cite{Stashans2003}, and ferroelectric modes~\cite{Edge2015, Gabay2017, Arce-Gamboa2018}.
    
The development of STO-based interfaces~\cite{Ohtomo2004, Pai2018a}, like the LaAlO$_3$/SrTiO$_3$ (LAO/STO) interface, has revived interest in the superconducting properties of STO, particularly following key experimental reports in the LAO/STO system~\cite{Reyren2007, Caviglia2008, Bell2009, Gariglio2009}. Superconductivity at this interface exhibits many of the same features as bulk STO, including the superconducting dome as well as characteristic temperature ($T_c \leq 0.4$ K) and magnetic field ($H_{c2} \sim 2$ kOe) scales, and carrier densities in the $10^{12-13}$ cm$^{-2}$ range, comparable to corresponding densities for bulk STO~\cite{Lin2013}. Quantum dots and nanowires created at the LAO/STO heterointerface exhibit electron pairing without superconductivity~\cite{Cheng2015}, which provides us with an independent measure of the pairing strength $E_P \sim 0.1 - 1 \, \text{meV}$. 
    
Ferroelastic domains are ubiquitous at the LAO/STO interface and their intersecton with the LAO/STO interface strongly influences the transport behavior~\cite{Honig2013, Kalisky2013, Frenkel2017}. The domains observed at the surface are of two kinds. First, ferroelastic domain structures naturally form bulk STO~\cite{Scott2012,Salje2013}, and some of these bulk domains intersect the surface. Second, there is evidence that electron density variations can nucleate ferroelastic domains at the LAO/STO interface. Specifically, piezoforce microscopy techniques shows that at room temperature electron-rich regions expand along the $c$-axis perpendicular to the interface~\cite{Bi2016}, which seeds the formation of $z$-oriented ferroelastic domains at low temperatures. Local probes have shown that these domain boundaries strongly influence carrier transport, both in the normal and superconducting state~\cite{Honig2013, Kalisky2013, Roy2017, Cheng2018}. 
    
\begin{figure*}
    \centering
    \includegraphics[width=\textwidth]{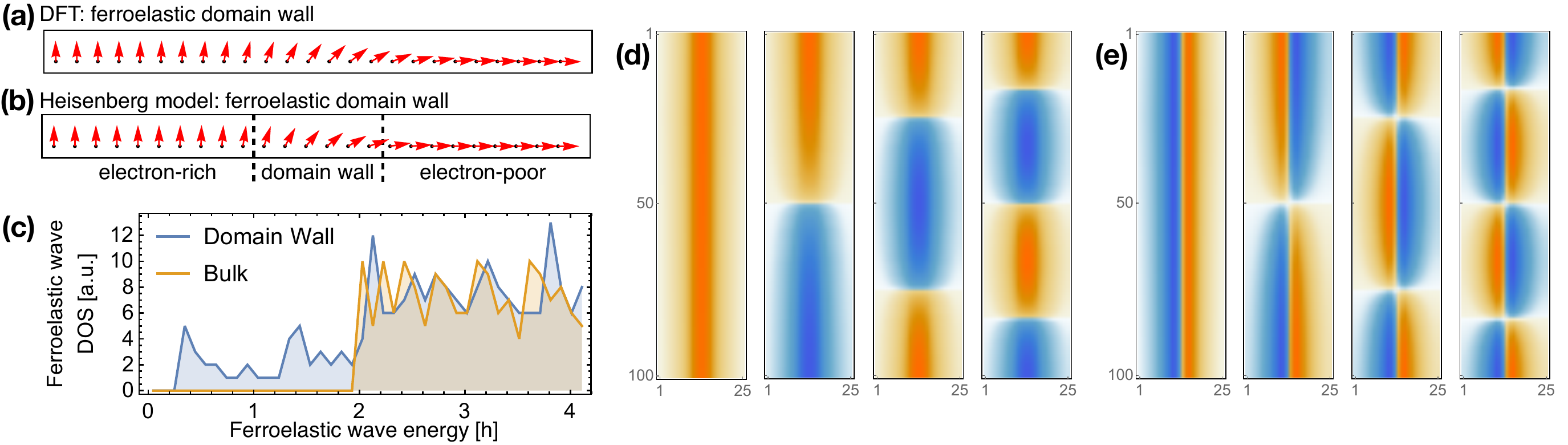}
    \caption{Ferroelastic waves in the presence of a ferroelastic domain wall. (a) Ferroelastic distortion as a function of position across a domain wall [arrows pointing up (right) correspond to $z-$ ($x-$) axis elongation] obtained from ab initio calculations. (b) Same as (a), but obtained from the Heisenberg model of Eq.~\eqref{eq:HFE}.
    (c) Density of states of ferroelastic waves with and without a domain wall. In the absence of a domain wall, the spectrum of ferroelastic waves has a gap of $2 |h|$, where $|h|$ is the magnitude of the locking field. This gap is significantly reduced in the presence of a domain wall (becoming 0.33 B for the case depicted). (d) Real-space mode profiles of the 4 lowest in energy ferroelastic waves. The mode weight is almost completely contained inside the domain wall that runs up-and-down through the center of each panel. (d) Real-space mode profiles of the 4 lowest in feroelastic waves associated with the second peak in the ferroelastic density of states, which appears at $E/B\sim 1.3$. These set of modes correspond to second transverse harmonic within the domain wall.  }
    \label{fig:spin-wave}
\end{figure*}

Recent experiments by Pai et al.~\cite{Pai2018} provide evidence for an intrinsic 1D nature of the superconducting state in LAO/STO.  In those experiments, a conductive atomic-force microscopy lithography technique~\cite{Cen2008,Cen2009,Brown2016} was used to define electron-rich channels with widths that varied systematically between 10~nm and 1~$\mu$m. These  channels showed a superconducting critical current that was independent of the width of the channel. Further, in experiments with multiple parallel channels, the critical current was found to be proportional to the number of conducting channels. 
 
The boundaries of the conductive channels are known to coincide with ferroelastic domain walls, which suggest that they play a role in 1D electron pairing. $z$-oriented ferroelastic domains form under the electron-rich (conducting) regions, while the electron poor (insulating) regions are associated with $x$- or $y$-oriented domains~\cite{Honig2013, Bi2016}. Ferroelastic domain walls form at the interface of the conducting and insulating regions, precisely where superconductivity seems to appear~\cite{Pai2018}. These type of domain walls, with a thickness of several lattice sites, have been investigated in first-principles calculations in Ref. \cite{Hellberg2019}.
    
Here we describe a model of electron pairing in which the coupling between electrons and ferroelasticity naturally leads to an attractive electron-electron interaction, with the strongest attractive interactions occurring along the ferroelastic domain walls. Specifically, we describe electron pair formation near a ferroelastic domain wall that forms between an electron-rich and an electron-poor region. We begin by extracting the typical ferroelastic domain wall width from first principles density functional theory calculations. Using a spin-wave-like analysis, we show that the ferroelastic domain walls host low-energy elastic modes. Next, by coupling the ferroelastic deformations to electron density, we show that ferroelastic deformations mediate attractive interactions between pairs of electrons. These interactions are found to be strongest when the two electrons are near the domain wall. Finally, we investigate electron pairing using a real-space analogy to the Cooper pair problem~\cite{Cooper1956}. We consider two electrons in the vicinity of the domain wall: (1) the electrons are restricted from entering the electron rich region due to the Pauli exclusion principle; (2) the electrons experience short-range repulsion; (3) the electrons experience long-range attraction as well as attraction to the domain wall that is mediated by ferroelastic distortions and is described at the level of the Born approximation in which electrons are treated as heavy particles. We find that electrons can indeed bind into (real-space) Cooper pairs. We conclude by commenting about the implications of our model to superconductivity in patterned and bulk LAO/STO heterointerfaces as well as possible generalizations to describe superconductivity in bulk STO.
    
\section{Modeling ferroelastic domain walls}
We begin by constructing a 2D model of ferroelasticity at the surface of the LAO/STO heterointerface. Our goal is specifically to model the fluctuations of domains induced by charge density at the LAO/STO interface as opposed to bulk domains in STO. First principles density functional calculations show that ferroelastic domain walls are extended objects with a size of roughly $3\,\text{nm}$~\cite{Hellberg2019}. 

The largest distortion in tetragonal STO is the rotation of the oxygen octahedra, which can be used to define a vector that rotates by 90 degrees across a domain wall~\cite{Schiaffino2017}. To characterize the domain wall, we take the data of Ref.~\cite{Hellberg2019} and compute the local rotation vector. For each Ti atom the local rotation vector is 
\begin{align}
    \vec{v}_i=\frac{1}{4}\sum_j \vec{\delta v}_j \times \vec{e}_j
\end{align}
where $i$ labels the Ti atom, the sum runs over its six neighboring oxygens $j$, $\vec{\delta v}_j$ is the displacement vector of the $j$th oxygen with respect to its ideal position, and $\vec{e}_j$ is the unit vector from the titanium atom to the ideal position of the $j$th oxygen. Thus in the bulk tetragonal structure, $\vec{v}_i$ gives the displacement of the planar oxygens from their ideal positions.  We plot $\vec{v}_i$ as a function of position across the domain wall in Fig.~\ref{fig:spin-wave}a. The length of this vector varies from $0.315$~\AA\ in the bulk regions to $0.353$~\AA\ in the center of the domain wall. 
These calculations were performed on a 17.5 nm supercell containing two identical ``head-to-tail" domain walls, but only one is shown; computational details are given in \cite{Hellberg2019}.

It is natural, therefore, to model the ferroelasticity using a Heisenberg model (as opposed to an Ising model that would have abrupt domain walls)
\begin{align}
    H_{\text{FE}}=-J \sum_{\langle ij \rangle} \sigma_i \cdot \sigma_j - \sum_i h_i \cdot \sigma_i,
    \label{eq:HFE}
\end{align}
where $\sigma_i$ are the spin-1/2 operators representing the ferroelastic distortions, $J$ represents the elastic modulus, and $h_i$ is an effective magnetic field that locks the ferroelasticity in the $z$-direction in the electron rich region and the $x$-direction in the electron poor region (see Fig.~\ref{fig:spin-wave}b). We justify the use of the locking field by noting that away from domain walls, ferroelastic domains are locked to the heterointerface surface and the crystallographic axis. The locking field has an important implication as it gives mass to the spin-wave (Goldstone) mode which would otherwise be massless.
    
We construct the ferroelastic-wave spectrum of $H_{\text{FE}}$ using mean-field theory + spin-wave fluctuations analysis. We begin by writing down a trial wave function for the spins
\begin{align}
    |\psi[\{\phi_i\}]\rangle=\prod_i \left[\phi_i |\downarrow\rangle_i + \sqrt{1-|\phi_i|^2} |\uparrow\rangle_i\right],
    \label{eq:trPsi}
\end{align}
where the optimization parameters is the set of complex numbers $\{\phi_i\}$. Next, we find the variational ground state by minimizing $\langle\psi[\{\phi_i\} |  H_{\text{FE}} |\psi[\{\phi_i\}]\rangle$ with respect to the $\phi_i$'s to obtain the mean field ground state defined by $\phi^{0}_i$. Finally, we expand in small fluctuations around the variational ground state $\phi_i \rightarrow \phi^{0}_i + \delta_i(t)$, and minimize the action  $\langle\psi[\{\phi_i\} | i\partial_t - H_{\text{FE}} |\psi[\{\phi_i\}]\rangle$ to obtain the ferroelastic-wave spectrum.
    
\begin{figure*}
    \centering
    \includegraphics[width=0.9\textwidth]{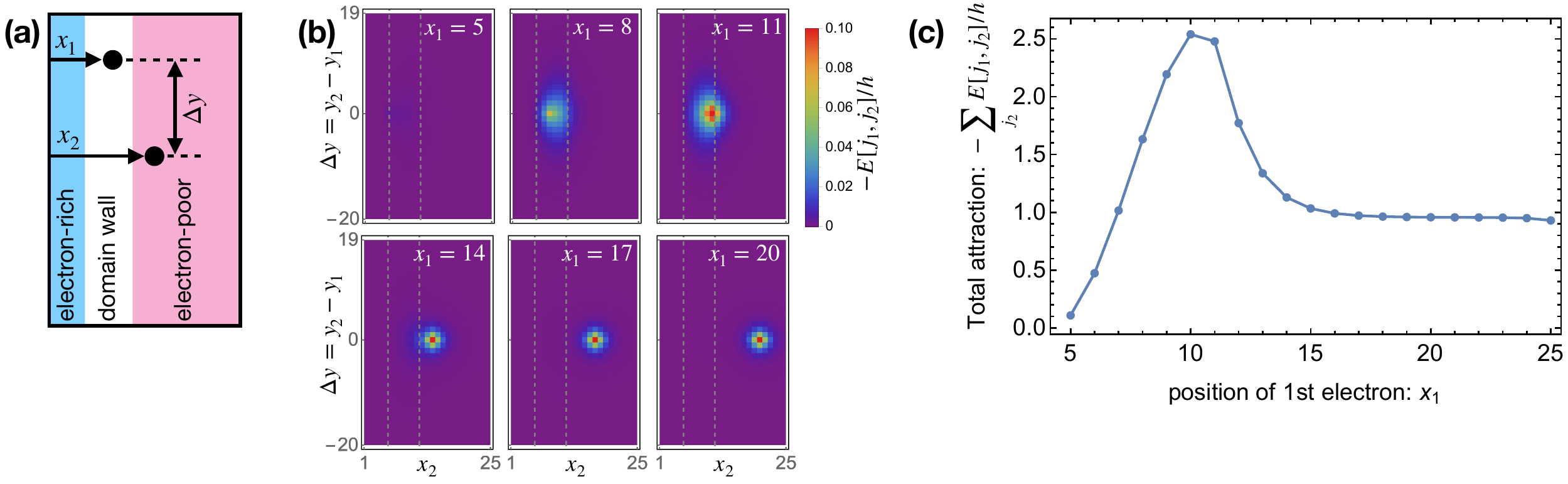}
    \caption{Electron-electron attraction mediated by ferroelastic deformations. (a) Setup of the calculations: two electrons are placed in the vicinity of a domain wall (at positions $j_1=\{x_1,y_1\}$ and $j_2=\{x_2,y_2\}$). The sum of the ferroelastic and interaction energy is minimized to obtain the effective electron-electron interaction energy. (b) Electron-electron interaction energy, mediated by ferroelastic deformations. The first electron is fixed at lattice site $x_1=\{5,8,11,\dots\}$ and the electron-electron interaction energy is plotted as a function of position of the 2nd electron. [In constructing this plot we subtracted off single-electron energy]. Interactions propagate further and are stronger when both electrons are in the vicinity of the domain wall. (c) Total electron-electron attraction energy as a function of the position of the 1st electron showing that the attraction is strongest when the 1st electron is inside the domain wall.   }
    \label{fig:energy}
\end{figure*}

Let us now consider a concrete example of a domain wall defined by the locking field
\begin{align}
    h(x,y)=\left\{ 
    \begin{array}{cc}
    \{0,0,1\} & x \leq 5\\
    \{1,0,0\} & x \geq 12
    \end{array}
    \right.,
\end{align}
where $1 \leq x \leq 25$, $1 \leq y \leq 100$, we set $J=1.5$, and we use open boundary conditions in the $x$-direction and periodic in the $y$-direction. The ground state ferroelastic (spin) configuration for this domain wall has elongation in the $z$-direction for $x \leq 5$ simulating an electron-rich region and in the $x$-direction for $x \geq 12$ simulating an electron-poor region (see Fig.~\ref{fig:spin-wave}b). Spin-wave theory tells us that away from a domain wall, the ferroelastic-wave spectrum has a gap of $2|h|$, where $|h|$ is the magnitude of the locking field (we use the bulk value of $|h|$ as the energy scale in the remainder of the manuscript). By introducing a domain wall as depicted in Fig.~\ref{fig:spin-wave}b, we find that the ferroelastic-wave spectrum acquires a set of modes inside the bulk gap (see Fig.~\ref{fig:spin-wave}c). The mode weight, $\delta_i$, of these low energy modes lies almost entirely inside the domain wall, see Fig.~\ref{fig:spin-wave}d and e. 
    
\section{Electron-electron attraction}
We investigate electron-electron interactions mediated by introducing a linear coupling between the electron density and ferroelasticity. This coupling is inspired by the experimental evidence that the electron density at the LAO/STO interface directly couples to an expansion of the crystal perpendicular to the surface~\cite{Bi2016}. Hence, we choose the interaction Hamiltonian
\begin{align}
    H_{\text{int}}=-\alpha \sum_{i,\sigma} n_{i,\sigma} S^z_i,
\end{align}
where $n_{i,\sigma}=c^\dagger_{i,\sigma}c_{i,\sigma}$ is the electron number operator and $\alpha$ is the coupling constant (we set $\alpha=h$ in the remainder of the manuscript).
    
To estimate the electron-electron interaction energy, we invoke the Born approximation and treat the electrons as heavy particles and ferroelastic waves as light particles. We begin by considering the effect of a single electron on the ferroelasticity. Within our model, an extra electron placed in an electron-rich domain has very little effect on the ferroelasticity as it is already full polarized. On the other hand an electron placed in an electron-poor domain results in a deformation of the ferroelasticity that heals on the over the length-scale $\xi_{\text{bulk}}=\sqrt{J/h}$. An electron placed in a domain wall results in the deformation of the ferroelasticity, mainly inside the domain wall, that heals over a length-scale $\xi_{\text{DW}} \approx \sqrt{w}$, where $w$ is the domain wall width. 

To compute electron-electron interaction, within the Born approximation, we need to find the energy $H_{\text{FE}}+H_{\text{int}}$ for each electron configuration (that is defined by the positions of the two electrons). Therefore, we place two electrons, of opposite spin, at positions $j_1=\{x_1,y_1\}$ and $j_2=\{x_2,y_2\}$ (see Fig~\ref{fig:energy}a) and use the trial wave function of Eq.~\ref{eq:trPsi} to minimize the energy of this configuration
\begin{align}
    E[j_1,j_2] = \langle \psi[\{\phi_i\} |  H_{\text{FE}} + H_{\text{int}}[j_1,j_2] |\psi[\{\phi_i\}]\rangle.
    \label{eq:E2e}
\end{align}
As the system is invariant with respect to displacements along the domain wall, the two-electron energy is described by three parameters: 
\begin{enumerate}
    \item $x_1$: the displacement of the first electron away from the center-line of the domain wall, 
    \item $x_2$:  the displacement of the first electron away from the center-line of the domain wall, 
    \item $\Delta y = y_2-y_1$: the displacement of the two electrons along the domain wall.
\end{enumerate} 

The energy in Eq.~\eqref{eq:E2e} is composed of the electron-electron interaction energy $E_2$ and the single-electron energy $E_1$
\begin{align}
E[j_1,j_2] = E_2[j_1,j_2]+E_1[j_1]+E_1[j_2].
\end{align}
$E_2$ describes the electron-electron attraction mediated by the ferroelasticity; $E_1$ describes the attraction of electrons to electron-rich region, which is also mediated by the ferroelasticity. 
We extract the two-electron energy $E_2$ from the computed energy $E[x_1,x_2,\Delta y]$ using the formula
$E_2[x_1,x_2,\Delta y] \approx E[x_1,x_2,\Delta y]-E[x_1,x_2,\Delta y\rightarrow L_y/2]$, where $L_y/2$ is the maximum value of $\Delta y$ for two electrons in a box of length $L_y$ along the $y$ directions. In Fig.~\ref{fig:energy}b we plot $E_2$. As $E_2$ is a function of three parameters ($x_1$, $x_2$ and $\Delta y$), we choose to fix $x_1$ (i.e. the position of the 1st electron) and vary $x_2$ and $\Delta y$ (i.e. the position of the 2nd electron) in each panel of Fig.~\ref{fig:energy}b. 

\begin{figure}
    \centering
    \includegraphics[width=\columnwidth]{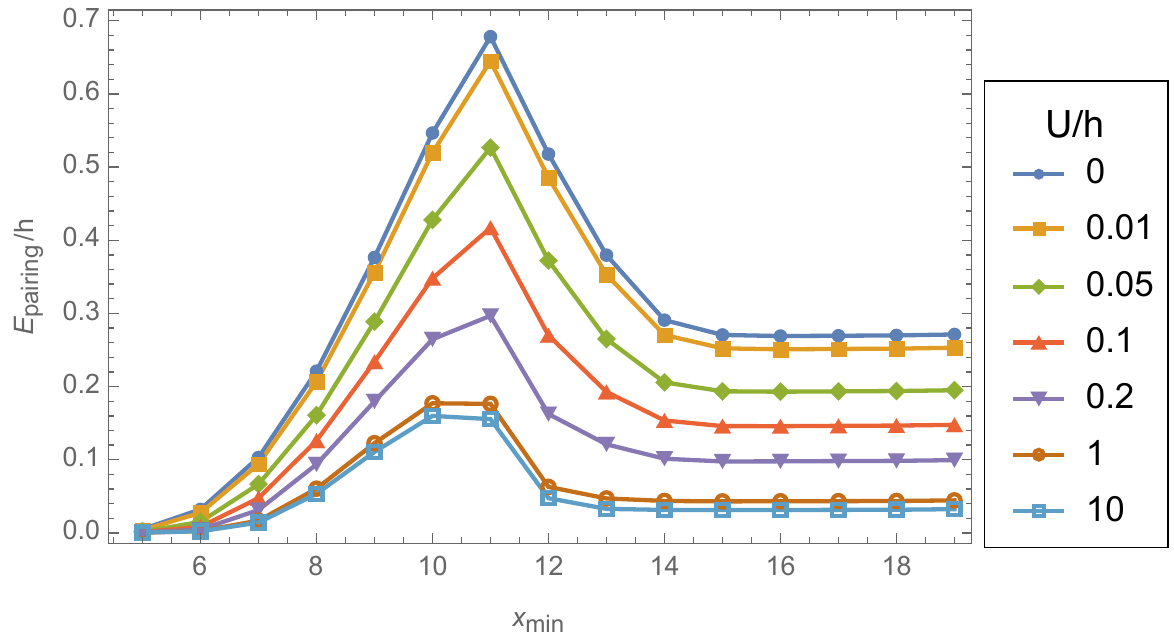}
    \caption{Binding energy of a real-space Cooper pair as a function of the barrier position $x_\text{min}$. The barrier prevents the occupation of sites to the left $x_\text{min}$ by the two electrons that are forming the pair. The pairing energy was computed for a number of values of the on-site repulsion $U$.}
    \label{fig:pairBindingEnergy}
\end{figure}

We find that $E_2[x_1,x_2,\Delta y]$ attains a minimum (indicative of electron-electron attraction) when the two electrons are near each other. The spatial extent of the ``dip'' around the minimum depends on the position of the first electron $x_1$. When the 1st electron is in the electron-poor region the dip has a smaller spatial extent than when the 1st electron is inside the domain wall. 

To quantify the size of this dip, we plot the integrated two-electron energy $E[j_1]=\sum_{j_2} E_2[j_1,j_2]$ as a function of the distance between the first electron and the domain wall. When the first electron is inside the electron rich region (on the left side of the domain wall) there is essentially no electron-electron attraction. This is due to the fact that ferroelasticity is already maximally deformed and the addition of one more electron has no effect. Similarly, the electron-electron attraction is also very weak when the first electron is located in the electron-poor region. This is a consequence of the fact that ferroelastic deformations induced by en electron in the electron-poor region have a very short range that is associated with the large gap to ferroelastic modes. Finally, when the first electron is in the middle of the domain wall, it induces long-range ferroelastic deformations (that propagate along the domain wall) and thus we observe strong electron-electron attraction.

\section{Real space Cooper-pair model}
In order to demonstrate that electrons can form pairs in the vicinity of a ferroelastic domain wall, we construct a Cooper-pair-like model in real (as opposed to momentum) space. Specifically, we concern ourselves with the motion of two  electrons with opposite spin. The motion of the two electrons is described by the Hubbard model
\begin{align}
    H_{\text{el}}=-t\sum_{\langle ij \rangle, \sigma} c^\dagger_{i,\sigma} c_{j,\sigma} + U \sum_{i} n_{i,\uparrow} n_{i,\downarrow} + \sum_{i,j} E[i,j] n_{i,\uparrow} n_{j,\downarrow}
    \label{eq:el}
\end{align}
with hopping amplitude $t$, on-site repulsive interaction $U$, and long-range electron-electron interaction mediated by ferroelasticity $E[i,j]$ from Eq.~\eqref{eq:E2e} (here we include both the one- and two-electron terms). 

To understand the contribution of ferroelastic domain walls to the strength of electron-electron pairing, we consider two cases: (1) a pair of electrons in the electron-poor bulk; (2) a pair of electrons in the vicinity of a ferroelastic domain wall. To be able to smoothly interpolate between these cases, in analogy to the Cooper problem, we supplement the electron Hamiltonian of Eq.~\eqref{eq:el} with the condition that the two select electrons must remain to the right side of a barrier located at $x=x_0$.

In Fig.~\ref{fig:pairBindingEnergy} we plot the electron pair binding energy as a function of $x_0$ for various values of the on-site repulsion $U$ (with $t=0.25 h$). As $x_0$ increase the electron pair is pushed into the middle of the domain wall, and we observe strong pairing. As we keep increasing $x_0$ the electrons are pushed out of the domain wall and into the electron poor region. Consequently, the pair binding energy decreases as we would expect from Fig.~\ref{fig:energy}. As we increase the on-site repulsive interaction $U$, we observe that the pair binding energy decreases. The pair binding energy decreases proportionately more in the electron-poor region as compared to the domain wall region, as the electron-electron interactions have a shorter range in the electron-poor region as compared at the domain wall. 

To summarize, we observe that ferroelastic domain walls enhance the electron pair binding energy and make electron-pairing more robust to local repulsive interactions. We believe that this enhancement helps to mediate superconductivity in the LAO/STO interface.

\section{Summary and outlook}
In summary, we have presented a scenario ascribing the mechanism of superconductivity in LAO/STO heterostructures to ferroelastic domain walls that form at the interface of electron-rich and electron-poor regions. Specifically, we built up a model that encompass our scenario. First, we modeled ferroelastic waves and showed that ferroelastic domain walls support low-energy modes analogous to capillary waves at the interface of two fluids. Second, we coupled our model of ferroelasticity to electron density and showed that electron-electron attraction is indeed strongest and also has the longest range in the vicinity of domain walls. Finally, we computed the electron-electron binding energy for a select pair of electrons, in analogy to the Cooper-pair problem, and showed that ferroelastic domain walls indeed enhance electron binding. 

The scenario that we present is consistent with available data on LAO/STO heterointerfaces. It naturally provides an explanation for the one-dimensional nature of superconductivity reported in Ref.~\cite{Pai2018}. It may also help to explain the origin of the superconducting dome that is observed as a function of the carrier density~\cite{Lin2013}: as we increase the electron density pairing first becomes stronger as the gap to ferroelastic modes decreases; however, as electron density become higher ferroelasticity becomes fully saturated in the $z$-direction resulting in the closing of the bulk superconducting gap as superconductivity is pushed to the edges of the electron puddle. 

We note that the presented scenario predicts that superconductivity should be strongly affected by strain fields which could move existing domain walls or introduce new domain walls.

\begin{acknowledgments}
We thank Anthony Tylan-Tyler for useful discussion and performing initial analysis of the related Ising model domain walls. DP and JL acknowledge support from NSF grant PHY-1913034. DP acknowledges support from the Charles E. Kaufman Foundation under grant KA2014-73919. CSH acknowledges support from the Office of the Secretary of Defense through the LUCI program and a grant of computer time from the DoD High Performance Computing Modernization Program.  JL acknowledges support from the Vannevar Bush Faculty Fellowship program sponsored by the Basic Research Office of the Assistant Secretary of Defense for Research and Engineering and funded by the Office of Naval Research through grant N00014-15-1-2847.  

\end{acknowledgments}

\bibliography{STO_SC_Theory_Simple}

\end{document}